\documentclass[conference]{IEEEtran}
\IEEEoverridecommandlockouts
\usepackage{cite}
\usepackage{amsmath,amssymb,amsfonts}
\usepackage{algorithmic}
\usepackage{graphicx}
\usepackage{textcomp}
\usepackage{xcolor}
\usepackage{multirow}
\usepackage{makecell}
\usepackage{booktabs}
\usepackage{threeparttable}
\usepackage{hyperref}
\def\BibTeX{{\rm B\kern-.05em{\sc i\kern-.025em b}\kern-.08em
    T\kern-.1667em\lower.7ex\hbox{E}\kern-.125emX}}
\begin{document}

\title{SaleNet: A low-power end-to-end CNN accelerator for sustained attention level evaluation using EEG\\
}

\author{\IEEEauthorblockN{
Chao Zhang\IEEEauthorrefmark{1}\IEEEauthorrefmark{2}, 
Zijian Tang\IEEEauthorrefmark{1}\IEEEauthorrefmark{2}, 
Taoming Guo\IEEEauthorrefmark{1}, 
Jiaxin Lei\IEEEauthorrefmark{1}\IEEEauthorrefmark{2}, 
Jiaxin Xiao\IEEEauthorrefmark{1}, 
Anhe Wang\IEEEauthorrefmark{3}, 
Shuo Bai\IEEEauthorrefmark{3}, 
Milin Zhang\IEEEauthorrefmark{1}\IEEEauthorrefmark{2}}
\IEEEauthorblockA{\IEEEauthorrefmark{1}Department of Electronic Engineering, Tsinghua University,\\
\IEEEauthorrefmark{2}Institute for Precision Medicine, Tsinghua University,\\
\IEEEauthorrefmark{3}State Key Laboratory of Biochemical Engineering, Institute of Process Engineering, Chinese Academy of Sciences\\
Corresponding author email: zhangmilin@tsinghua.edu.cn}}

% \author{\IEEEauthorblockN{
% \thanks{\ } 
% \\
% \\
% }
% \IEEEauthorblockA{
% \\
% \\
% \\
% }}

\maketitle

\begin{abstract}
This paper proposes SaleNet - an end-to-end convolutional neural network (CNN) for sustained attention level evaluation using prefrontal electroencephalogram (EEG). 
A bias-driven pruning method is proposed together with group convolution, global average pooling (GAP), near-zero pruning, weight clustering and quantization for the model compression, achieving a total compression ratio of 183.11x.
The compressed SaleNet obtains a state-of-the-art subject-independent sustained attention level classification accuracy of 84.2\% on the recorded 6-subject EEG database in this work.
The SaleNet is implemented on a Artix-7 FPGA with a competitive power consumption of 0.11 W and an energy-efficiency of 8.19 GOps/W.
\end{abstract}


\section{Introduction}\label{intro}

% BMI and Sale
Brain-machine interfaces (BMIs) aim to establish an efficient communication channel between the human brain and the external world \cite{tang2020nature}. 
Sustained attention level evaluation is one of the most crucial scenarios of BMI. It can benefit ubiquitous scenes such as the diagnosis of attention deficit hyperactivity disorder (ADHD) \cite{barkley1997behavioral} and the interaction efficiency in classroom \cite{ko2017sustained}.

% Related Sale work 
EEG is one of the popular bio-markers in BMI or brain-computer interface (BCI) applications.
A number of literature has explored the relationship between EEG and human mental status \cite{lin2014wireless,Peng-2021,dataset-2021}. Both the time domain feature such as mobility \cite{dataset-2021}, and the frequency domain feature such as band energy \cite{lin2014wireless} were found to be related with sustained attention levels. 
However, the previous algorithms and systems are still designed based on handcraft features. 
Holding the insights that EEG is related with attention levels, an end-to-end neural network (NN) based scheme is expected to achieve better attention level classification accuracy.
In addition, although \cite{lin2014wireless} implemented the algorithm into mobile device, there is still a lack of low power hardware solution for long-term and real-time sustained attention level monitoring.

% Related CNN accelerator work
As NN based models achieved brilliant success in fields such like biomedical signal processing, the design of NN accelerators attracted more and more attention of researchers \cite{sze2017efficient,Yoo2020iscas-emotion-cnn, sahani2021tbiocas-seizure, lu2021tcasI-ECG}.
\cite{Yoo2020iscas-emotion-cnn} proposed BioCNN for the EEG based emotion recognition on field-programmable gate array (FPGA), which held a competitive low power consumption of 0.15 W. 
However, it utilized a traditional scheme of separate feature extracting and classifying modules. 
The feature extractor was offline while the CNN was only exploited as a classifier, which made it unsuitable to develop a portable BMI.
\cite{sahani2021tbiocas-seizure} and \cite{lu2021tcasI-ECG} developed end-to-end CNNs for EEG based seizure detection and electrocardiography (ECG) based beats classification, respectively. 
\cite{sahani2021tbiocas-seizure} achieved a near-100\% seizure detection accuracy on hardware, 
but the upper-1W power consumption was still an issue in clinical practice.
The light 1-d CNN in \cite{lu2021tcasI-ECG} obtained a high throughput efficiency as well as a high beats classification accuracy. However, it was a challenge to handle more difficult tasks in biomedical signal processing using the model with only 11k weights.

% This work
In this work, we proposes an end-to-end CNN architecture for the sustained attention level evaluation, namely SaleNet.
The contributions of this work are listed as follows:
\begin{itemize}
\item [a)]
We proposes a bias-driven pruning method for the model compression. A compression flow including group convolution, global average pooling (GAP), near-zero pruning, bias-driven pruning, weight clustering and quantization achieves 183.11x model compression ratio.
\item [b)]
The SaleNet achieves a subject-independent accuracy of 84.2\% on the binary attention level classification task, which aligns with the state-of-the-art.
\item [c)]
The SaleNet is implemented on FPGA with a competitive power consumption of 0.11 W and an energy efficiency of 8.19 GOps/W.
\end{itemize}

The rest of the paper is organized as follows.
Section II introduces the algorithm and hardware design of the proposed CNN architecture for low cost sustained attention level evaluation.
Section III illustrates the experimental results, while Section IV concludes the entire work.

\section{Algorithm and hardware design}\label{sys}
\subsection{Architecture of the SaleNet}
\begin{figure}[thbp] %H为当前位置，!htb为忽略美学标准，htbp为浮动图形
    \centering %图片中
        \includegraphics[width=0.45\textwidth]{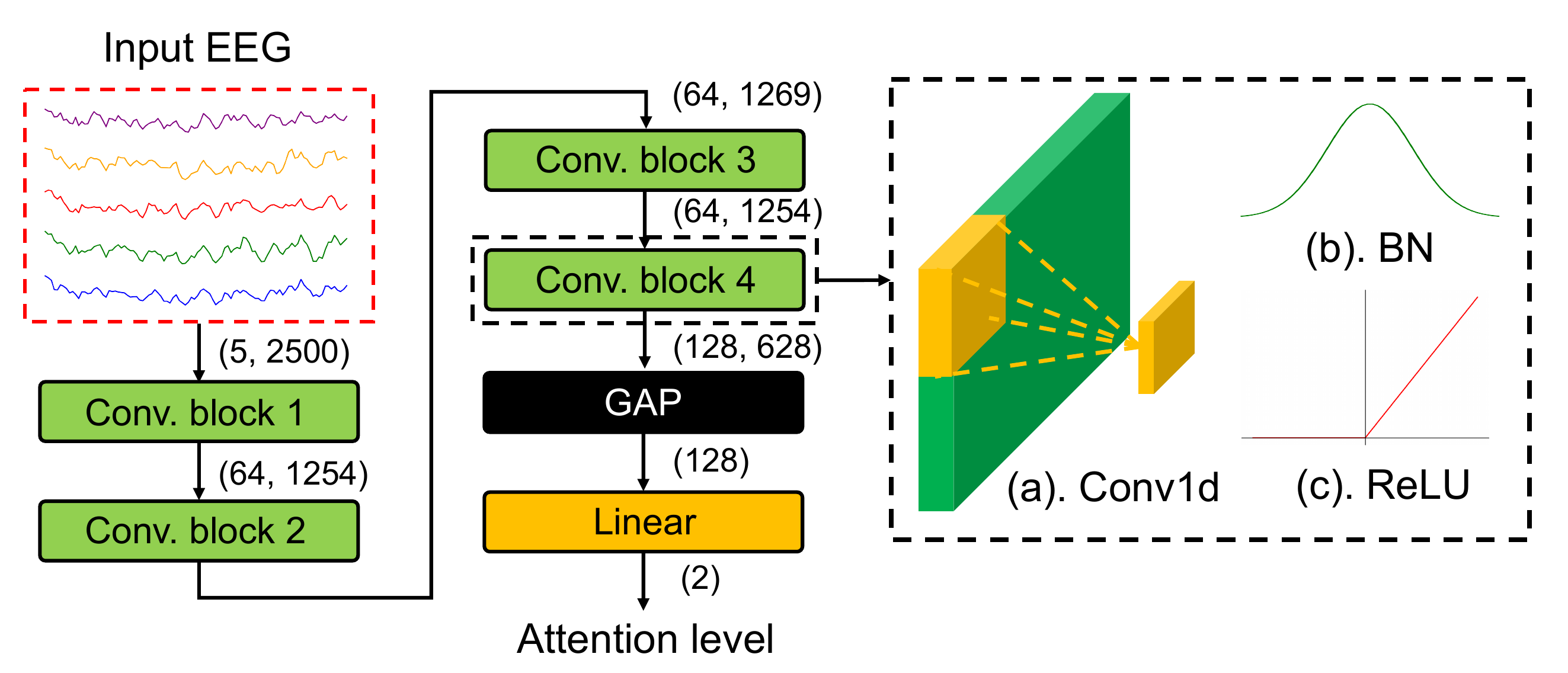} %插入图片，[]中设置图片大小，{}中是图片文件名
        \caption{Design of the SaleNet} %最终文档中希望显示的图片标题
    \label{fig:salenet} %用于文内引用的标签
\end{figure}

The proposed SaleNet is shown in Fig. \ref{fig:salenet}, which consists of four convolutional blocks, a GAP layer as well as a linear layer.
Each convolutional block includes a 1-d convolutional layer, a 1-d batch normalization (BN) layer and a rectified linear unit (ReLU) layer.
The kernel size of each convolutional layer is 16.
In order to compress the CNN model, the group convolution as shown in Equ. \ref{eq:group-conv} is applied in the last three convolutional layers with group number of 8, 8 and 16.
\begin{equation}
\label{eq:group-conv}
\begin{aligned}
out(i) = bias(i) + \sum_{k=s}^{e}w(i,k)*in(k), \, i \in [0,C_{out})
\end{aligned}
\end{equation}
The boundary points $s$ and $e$ in group convolution are given by Equ. \ref{eq:group-s} and \ref{eq:group-e}, where $g$ is the group number.
\begin{equation}
\label{eq:group-s}
\begin{aligned}
s = i \, // \, \frac{C_{out}}{g} \times \frac{C_{in}}{g}
\end{aligned}
\end{equation}
\begin{equation}
\label{eq:group-e}
\begin{aligned}
e = s + \frac{C_{in}}{g} - 1
\end{aligned}
\end{equation}
The $C_{out}$ and $C_{in}$ denote the channel numbers of the output map and the input map, respectively.
The normal convolutional layer is equivalent with the condition of $g=1$ in Equ. \ref{eq:group-conv}.
Since the group convolution treats the input feature map as individual groups, both the weight number and operation number of the convolutional layers can be reduced by $g$ times.

The linear layer in CNN usually requires a huge memory and computational load.
In SaleNet, the GAP is applied to avoid the large linear layer.
It can reduce both the weight number and operation number of the linear layers to $1/p$, where $p$ is the pooling size.

\subsection{Near-zero pruning and negative pruning}
ReLU makes a non-linear reflection from negative value to zero as shown in Equ. \ref{eq:relu},
\begin{equation}
\label{eq:relu}
\begin{aligned}
ReLU(x) = max(x, 0)
\end{aligned}
\end{equation}
which significantly increases the sparsity of the data flow in the forward propagation and makes it possible to reduce the model size as well as the computational load using pruning methods such as zero-pruning.
This study recruits two pruning strategies, namely near-zero pruning and bias-driven pruning.
Since the GAP dramatically reduces the size of the linear layer, the convolutional layers dominate the parameter and the operation numbers of the SaleNet, with contributions of 97.11\% and 97.82\%, respectively.
As a result, the pruning methods are designed for the weights of the convolutional layers.
\subsubsection{Near-zero pruning}
In the classification task, the accuracy depends only on the comparing results rather than the exact values of the output from the final linear layer.
It is practicable to remove several weights with small absolute values, which are usually not related to the model accuracy. 

\subsubsection{Bias-driven pruning}
\begin{figure}[thbp] %H为当前位置，!htb为忽略美学标准，htbp为浮动图形
    \centering 
        \includegraphics[width=0.4\textwidth]{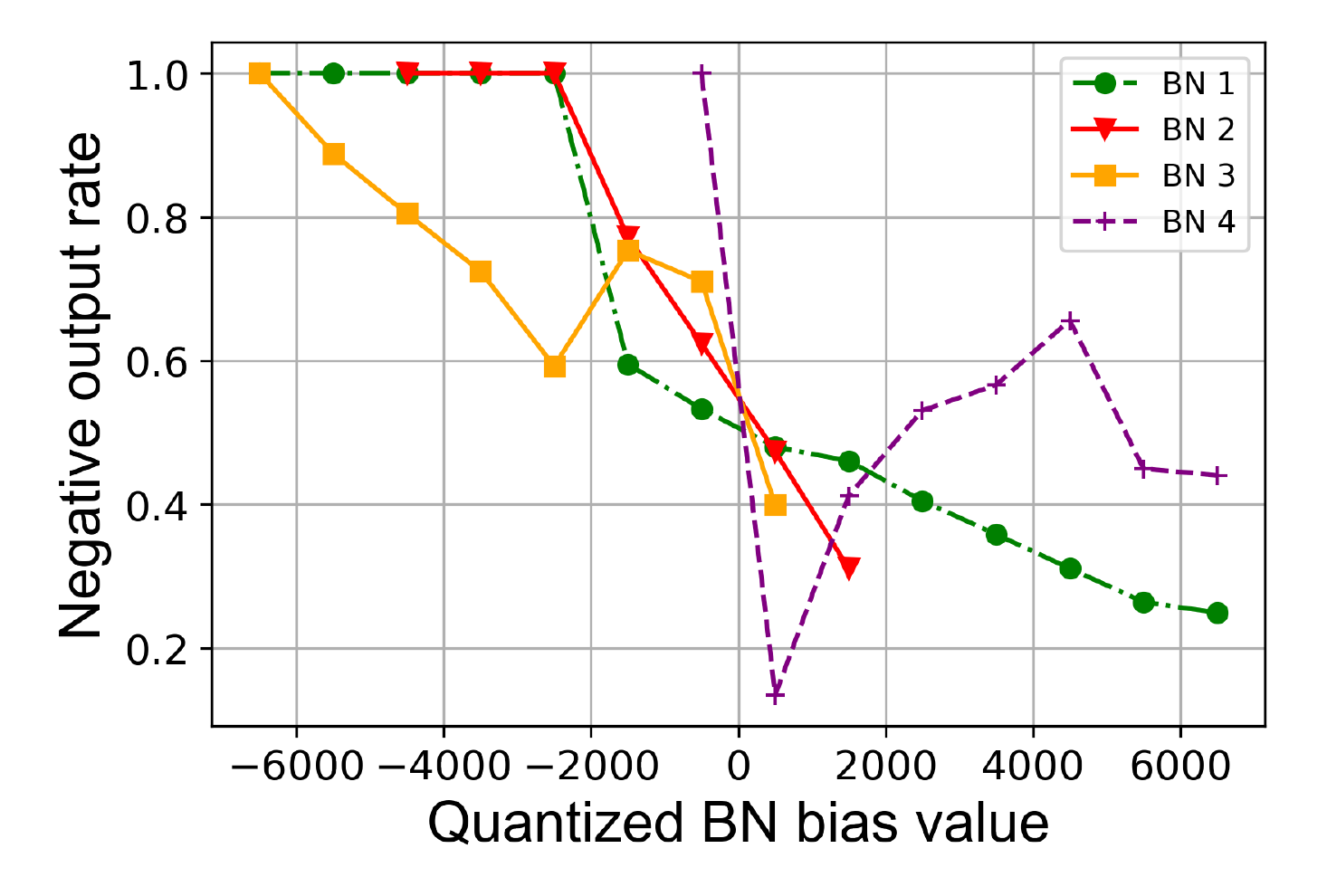} 
        \caption{The relationship between the BN bias and the negative rate of the BN output} 
    \label{fig:negrate} 
\end{figure}

This paper proposes an efficient pruning method based on the bias of the BN layers.
Since there is always a ReLU layer behind each BN layer, the negative output of the BN layer makes no sense to the inference result of the network.
As shown in Fig. \ref{fig:negrate}, the negative rate of the BN output is strongly related with the BN bias in most BN layers.
Thus the BN bias can be a indicator to decide whether the corresponding convolution and BN should be executed.
In SaleNet, we set three different thresholds for the first three BN layers.
For each channel of the BN output, when the BN bias is smaller than the threshold, the convolutional weights of the channel are pruned.

\subsection{Weight clustering and quantization}
Low bit quantization can reduce both the memory requirement and the operation width of the model, which will further benefit the power consumption and the inference latency. 
After network pruning, we exploits weight clustering to reduce the bit width of the weights of the convolutional layers.
The K-means clustering is applied to the weights of each convolutional layer to achieve a sub-8bit convolutional model.

For the quantization phase, there is a compromise between the model size and the classification accuracy. 
While keeping the classification accuracy above 80\% for the attention evaluation, we reduce the model width as low as possible.

\subsection{Hardware implementation}
\subsubsection{Process engine}
As demonstrated in Equ. \ref{eq:group-conv}, the 1-d convolution is equivalent with several inner product operations.
The BN layer in SaleNet is illustrated in Equ. \ref{eq:BN}.
\begin{equation}
\label{eq:BN}
\begin{aligned}
out_{BN} = \frac{x-E[x]_{BN}}{\sqrt{Var[x]_{BN}+10^{-5}}} \times \gamma + \beta
\end{aligned}
\end{equation}
The $\gamma$ and $\beta$ are two learnable parameters. 
The $E[x]_{BN}$ and $Var[x]_{BN}$ denote the mean and variance of each channel in a mini-batch, which are calculated and stored before the inference phase.

A process engine (PE) architecture according to Equ. \ref{eq:PE} is designed for both the 1-d convolutional block and the linear layer.
\begin{equation}
\label{eq:PE}
\begin{aligned}
out_{PE} = (\sum_{i=0}^{127}x[i]w[i]+b) \times w_{BN} + \beta 
\end{aligned}
\end{equation}
When utilized in convolutional block, the $w$ denotes the weight of the convolutional layer, while the $b$ and the $w_{BN}$ are calculated by Equ. \ref{eq:b} and \ref{eq:bn_w},
\begin{equation}
\label{eq:b}
\begin{aligned}
b = bias - E_{BN}
\end{aligned}
\end{equation}
\begin{equation}
\label{eq:bn_w}
\begin{aligned}
w_{BN} = \frac{\gamma}{\sqrt{Var_{BN}+10^{-5}}} 
\end{aligned}
\end{equation}
where the $bias$ denotes the convolutional layer bias.
When the PEs are recruited for the inference of linear layer, the $w$ and $b$ in Equ. \ref{eq:PE} denote the weight and bias of the linear layer, while the $w_{BN}$ and $\beta$ are set to 1 and 0, respectively.

\subsubsection{Hardware architecture}
\begin{figure}[thbp] %H为当前位置，!htb为忽略美学标准，htbp为浮动图形
    \centering %图片中
        \includegraphics[width=0.48\textwidth]{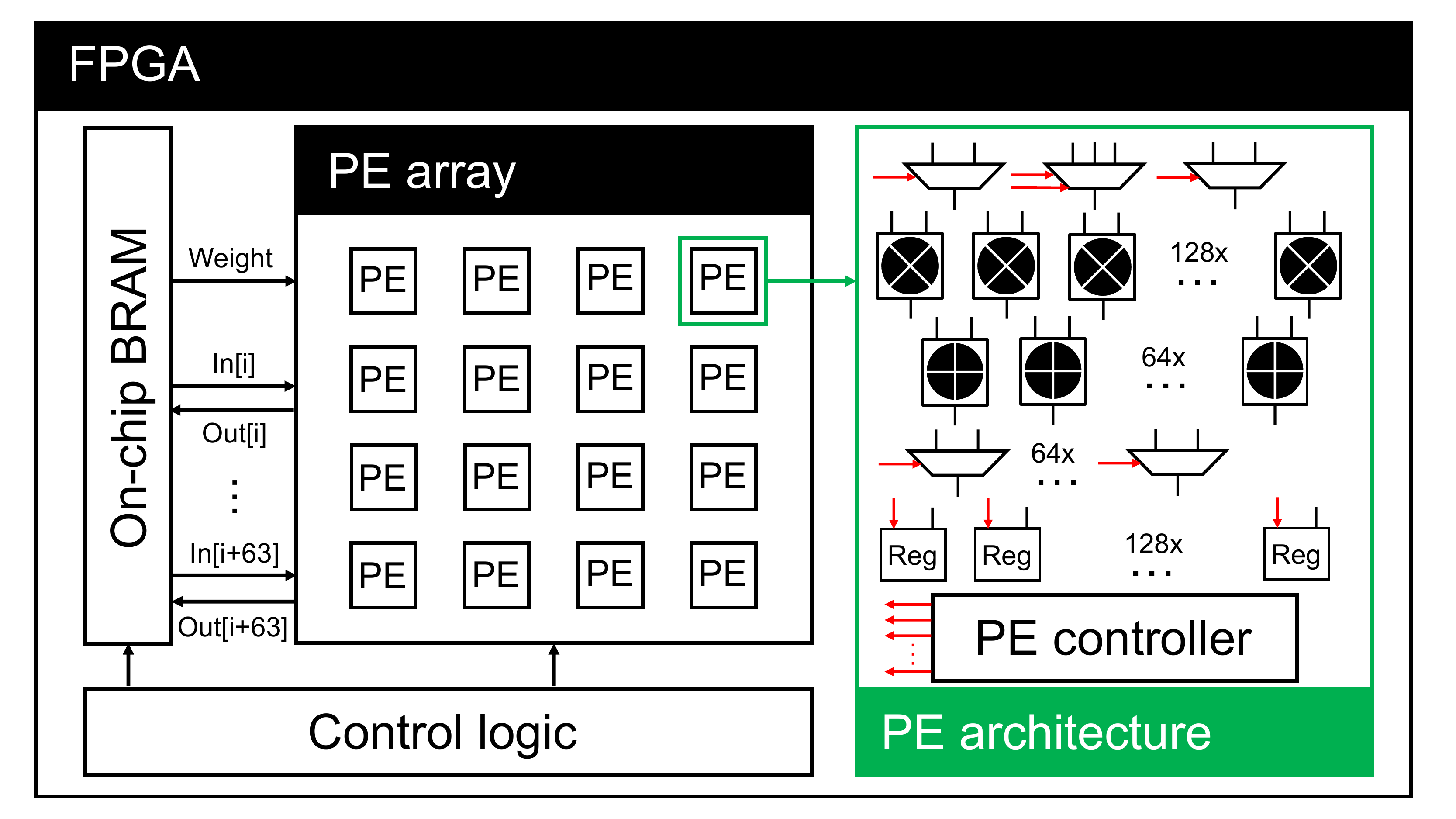} %插入图片，[]中设置图片大小，{}中是图片文件名
        \caption{Design of the hardware architecture} %最终文档中希望显示的图片标题
    \label{fig:fpga} %用于文内引用的标签
\end{figure}
The hardware architecture on FPGA is illustrated in Fig. \ref{fig:fpga}.
The weights of the SaleNet and the output of the first three convolutional blocks are stored in the on-chip block random access memory (BRAM).
Totally 16 PEs are implemented, each occupying 128 multipliers, 64 adders and 128 registers.
According to the network configuration, the inference of the four convolutional blocks needs PE cycles of 5016, 5076, 5016 and 5024, respectively.
Thanks to the GAP layer, the inference of the linear layer only needs one PE cycle.

Since the large number of the convolutional kernel channels and the long range of the input EEG signals, the output map of each convolutional block in SaleNet occupies more than 1Mb memory.
It is necessary to apply a memory re-use scheme for the output maps of the convolutional blocks.
To this end, the output maps of the first three convolutional blocks occupy the same space of the on-chip BRAM. According to the co-design of the group number and PE number, when a data slice is loaded into the PEs, the PE results can be directly stored to the corresponding address without affecting the subsequent calculations.
The output map of the last convolutional block is utilized directly for the GAP output.
For leveraging the latency and the power consumption, the data loading from the BRAM is completed in a fast clock domain of 50 MHz, while the PEs are executed in a slow clock domain of 10 MHz.

\section{Experimental Results}\label{exp}
\subsection{Data acquisition}
Six subjects were recruited for the EEG acquisition of this study. 
A number clicking task was designed to label the sustained attention levels of the subjects.
There was a 7 $\times$ 7 random number matrix on a screen. 
The subjects were asked to click number from 1 to an arbitrarily large number.
Once the subjects clicked the target number, the number matrix was refreshed and the target number increased by 1.
Each subject was asked to complete a 12-minute trial.
Each trial consists of four 3-minute phases. In the first and the third phases, the subjects were asked to relax as much as possible, while the subjects were asked to complete the number clicking task in the second and the fourth phases.
The data collected from the number clicking phases were labeled as high attention levels.
Seven channels prefrontal EEG of each subject was recorded in the designed 12-minute trial.
The EEG acquisition device is a module based on the neural signal recorder in \cite{luo2019low}.

\subsection{Sustained attention level evaluation}
During the evaluation, we randomly selected one subject as the test set and the other five subjects as the train set.
After training with a batch size of 16, the SaleNet achieved a subject-independent (SI) accuracy of 89.8\%.
For the requirement of real-time processing, we fixed the running mean and variance of the BN layers and evaluated the SaleNet using a batch size of 1, still obtaining an high SI accuracy of 85.9\%.
A comparison between the proposed SaleNet and several state-of-the-art attention evaluation works was shown in Table \ref{tb:SOTA-att}.
Compared to other works, the SaleNet provided an end-to-end CNN solution on FPGA with high SI accuracy, which is more possible to be integrated into portable BMIs.
\begin{table}[!ht]
\caption{Comparison with the state-of-the-art attention evaluation systems}
\label{tb:SOTA-att}
\begin{center}
\scriptsize
\begin{threeparttable}
\scalebox{1.0}{
\begin{tabular}{|c|c|c|c|c|}
\hline
& Lin \cite{lin2014wireless} & Delvigne \cite{dataset-2021} & Peng \cite{Peng-2021} & \textbf{This work} \\
\hline
Publication & TBioCAS & TCAS VT & T Haptics & \textbf{ISCAS} \\
\hline
Year & 2014 & 2021 & 2021 & \textbf{2022} \\
\hline
Experiment & Driving & Ball tracking & \makecell[c]{Fingertip\\forcing} & \textbf{\makecell[c]{Number\\clicking}} \\
\hline
Sample rate & 256Hz & 500Hz & 1000Hz & \textbf{250Hz}\\
\hline
\#EEG channel & 4 & 32 & 64 & \textbf{5}\\
\hline
Method & FFT+SVR & \makecell[c]{Temporal \\feature \\+ GNN}  & \makecell[c]{ERP \\+ PSD analysis} & \textbf{\makecell[c]{End-to-end \\CNN}}\\
\hline
Result & \makecell[c]{0.124 \\RMSE\tnote{a}}  & 72.4\% acc. & \makecell[c]{Highlighting \\ $\alpha$-band power}  & \textbf{84.2\% acc.} \\
\hline
Hardware & \makecell[c]{Mobile\\device} & None & None & \textbf{FPGA} \\
\hline
\end{tabular}}
 \begin{tablenotes}
        \footnotesize
        % \item[a] Support vector regression  
        % \item[b] Graph neural network
        \item[a] Root mean square error
        % \item[d] Event-related potential
      \end{tablenotes}
    \end{threeparttable}
\end{center}
\end{table}

\subsection{Compression of the SaleNet}
\subsubsection{Group convolution and GAP}
The group convolution and GAP compress the model size by 13.9x without significant evaluation accuracy deterioration as shown in Table \ref{tb:group-conv-gap}.
It should be highlighted that the GAP layer not only reduced the model size, but also suppressed the over-fitting issue to make sure the SaleNet achieve a high evaluation accuracy.
\begin{table}[!ht]
\caption{Results of the group convolution and GAP}
\label{tb:group-conv-gap}
\begin{center}
\scriptsize
\begin{threeparttable}
\scalebox{1.0}{
\begin{tabular}{c|c|c|c|c}
\hline\hline
Model & Parameters & Operations & \makecell[c]{Compression\\ratio} & Accuracy\tnote{*} \\
\hline
\hline
\makecell[c]{baseline \\ (4 layer CNN)} & 428.99k & 510.42M & 1x & 0.71\\
\hline
\makecell[c]{baseline + \\ GAP} & 268.48k & 510.26M & 1.6x & 0.89 \\
\hline
\makecell[c]{baseline + \\ group conv.} & 191.43k & 66.72M & 2.2x & 0.72\\
\hline
\makecell[c]{SaleNet \\ (baseline + \\ group conv. + \\ GAP)} & 30.91k & 66.56M & 13.9x & 0.89\\
\hline\hline
\end{tabular}}
 \begin{tablenotes}
        \footnotesize
        \item[*] Best accuracy in the first 100 epochs with a batch size of 16
      \end{tablenotes}
    \end{threeparttable}
\end{center}
\end{table}

\subsubsection{Pruning and quantization}
\begin{figure}[thbp] %H为当前位置，!htb为忽略美学标准，htbp为浮动图形
    \centering %图片中
        \includegraphics[width=0.48\textwidth]{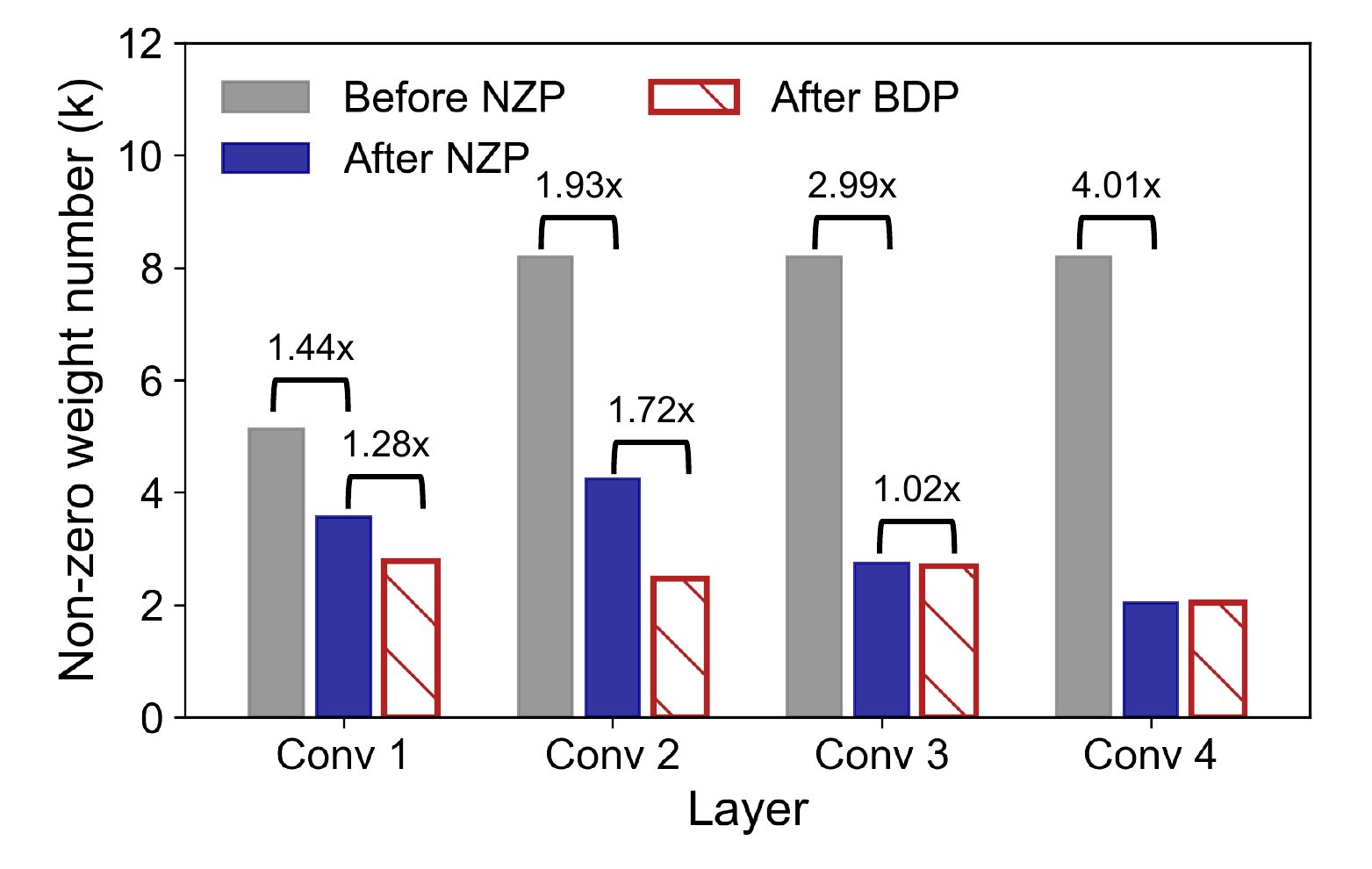} %插入图片，[]中设置图片大小，{}中是图片文件名
        \caption{Result of the weight pruning} %最终文档中希望显示的图片标题
    \label{fig:pruning} %用于文内引用的标签
\end{figure}

The results of the near-zero pruning (NZP) and the bias-driven pruning (BDP) were shown in Fig. \ref{fig:pruning}.
The near-zero pruning thresholds were set as 0.017, 0.014, 0.015 and 0.014 for the four convolutional layers. 
An average compression ratio of 2.36x was achieved.
According to the analysis in Fig. \ref{fig:negrate}, the bias-driven pruning was utilized on the first three convolutional blocks, with thresholds of -0.061, -0.046 and -0.183, respectively.
Results showed the proposed bias-driven pruning could further reduce 2.62k parameters after the near-zero pruning, corresponding a compression ratio of 1.26x.
There was no accuracy deterioration in the pruning step.

After the weight clustering and the quantization, the bit width of the convolutional weights, the convolutional bias, the BN weights, the BN bias, the linear weights and the linear bias were set to 7, 8, 16, 14, 8 and 11, respectively. Comparing to the standard 32 bit representation of the non-zero network parameters, this step contributed a compression ratio of 4.43x. 
The accuracy dropped from 85.9\% to 84.2\% during the quantization step, which was still reliable for the attention evaluation application. 

Comparing to the baseline model in Table \ref{tb:group-conv-gap}, the group convolution, the GAP, the near-zero pruning, the bias-driven pruning, the weight clustering together with the quantization contributed a total compression ratio of 183.11x.

% weight clustering: 0.859

% data: 1824, -2328. 13bit

% conv1.w: 34, -36  7bit
% conv2.w: 19, -19
% conv3.w: 15 -13
% conv4.w: 9 -10

% conv1.b: 1, -1.  8bit
% conv2.b: 11, -4
% conv3.b: 12, -5
% conv4.b: 13, -14
% E1: 0, 0
% E2: 64, -57
% E3: 15, -39
% E4: 8, -13

% bn1.w: 18040, 3479. 16bit
% bn2.w: 15369, 2073. 
% bn3.w: 12821, 991
% bn4.w: 28520, 1411
% Var1: 77910, 7006
% Var2: 6032, 0
% Var3: 3136, 0
% Var4: 266, 0

% bn1.b: 6988, -6433. 14bit
% bn2.b: 3903, -4400
% bn3.b: 1910, -6443
% bn4.b: 7049, -1

% linear.w: 67, -66  8bit
% linear.b: 885, -318. 11bit

\subsection{Hardware implementation}
The SaleNet consumed 65731 LUTs, 35199 FFs, 736 DSPs and 68 BRAMs on a Xilinx Artix-7 FPGA with the tool of Vivado 2017.4. Table \ref{state-of-the-art} showed a comparison between the SaleNet and several the state-of-the-art FPGA solutions in physiological signal processing.
Since the sustained attention level evaluation task is not sensitive to sub-second level delay, the SaleNet made a compromise between latency and power consumption.
Compared to other solutions, the SaleNet holds a competitive design power consumption of 0.11 W and an energy-efficiency of 8.19 GOps/W.
\begin{table}[!ht]
\caption{Comparison with the state-of-the-art FPGA solutions}
\label{state-of-the-art}
\begin{center}
\scriptsize
\begin{threeparttable}
\scalebox{1.0}{
\begin{tabular}{|c|c|c|c|c|}
\hline
& Gonzalez \cite{Yoo2020iscas-emotion-cnn} & Sahani \cite{sahani2021tbiocas-seizure} & Lu \cite{lu2021tcasI-ECG} & \textbf{This work} \\
\hline
Publication & ISCAS 20 & TBioCAS 21 & TCAS I 21 & \textbf{ISCAS 22} \\
\hline
Signal & EEG & EEG & ECG & \textbf{EEG} \\
\hline
Task & \makecell[c]{Emotion\\recognition} & \makecell[c]{Seizure\\detection} & \makecell[c]{Beats\\classification} & \textbf{\makecell[c]{Attention\\Evaluation}}\\
\hline
Sample rate & 128Hz & 256Hz & 360Hz & \textbf{250Hz} \\
\hline
\#Channel & 14 & 18 & 1 & \textbf{5} \\
\hline
Method & FE\tnote{*}+CNN & CNN & CNN & \textbf{CNN} \\
\hline
End-to-end & No & Yes & Yes & \textbf{Yes} \\
% \hline
% Clock & 100M & 86.73M & 200M & \textbf{50M/10M} \\
\hline
Precision & 16FD & 16FD & 16FD & \textbf{7-16FD} \\
% \hline
% Latency & $\leq$ 1ms & 30ms & - & \textbf{74ms} \\
\hline
Throughput & 1.65GOps & - & 25.7GOps & \textbf{0.90GOps} \\
\hline
Power & 0.15W & 1.59W & - & \textbf{0.11W} \\
\hline
\makecell[c]{Energy \\efficiency} & 11GOps/W & - & - & \textbf{8.19GOps/W} \\
\hline
Board & Spartan-6 & Virtex-5 & ZC706 & \textbf{Artix-7} \\
\hline
CNN size & - & - & 1.028MOPs & \textbf{66.56MOPs} \\
\hline
Parameter & - & - & 11.07k & \textbf{30.91k} \\
\hline
\end{tabular}}
 \begin{tablenotes}
        \footnotesize
        % \item[*] Estimated
        \item[*] Feature extractor
      \end{tablenotes}
    \end{threeparttable}
\end{center}
\end{table}

\section{Conclusion}
This paper proposes SaleNet - an end-to-end CNN for sustained attention level evaluation using 5-channel prefrontal EEG.
The SaleNet is compressed by 183.11x using the group convolution, the GAP, the near-zero pruning, the weight cluster, the quantization and the proposed bias-driven pruning.
The compressed SaleNet achieves a state-of-the-art subject-independent sustained attention level classification accuracy of 84.2\% .
As far as the authors know, the proposed SaleNet is the first end-to-end CNN solution on FPGA for the sustained attention level evaluation, which consumes a competitive design power of 0.11 W and holds an energy efficiency of 8.19 GOps/W.

\section*{Acknowledgment}
This work is supported in part by the National Key Research and Development Program of China (No.2018YFB220200*), in part by the Natural Science Foundation of China through grant 92164202. in part by the Beijing Innovation Center for Future Chip, in part by the Beijing National Research Center for Information Science and Technology.

\renewcommand\refname{Reference}
\bibliographystyle{IEEEtran}
\bibliography{cnn_fpga}

\end{document}